\documentstyle[aps,prl,multicol,epsf]{revtex}
\begin{document}

\title{\bf Extreme Value Statistics of Hierarchically Correlated
Variables: Deviation from Gumbel Statistics and Anomalous Persistence}
\author{D.~S.~Dean$^{1}$ and Satya N. Majumdar$^{1,2}$}
\address{$^1$CNRS, IRSAMC, Laboratoire de Physique Quantique,
Universite' Paul Sabatier, 31062 Toulouse, France} \address{$^2$Tata
Institute of Fundamental Research, Homi Bhabba Road, Mumbai-400005,
India}

\maketitle

\begin{abstract}{ We study analytically the distribution of the minimum 
of a set of hierarchically correlated random variables $E_1$, $E_2$,
$\dots$, $E_N$ where $E_i$ represents the energy of the $i$-th path of
a directed polymer on a Cayley tree.  If the variables were
uncorrelated, the minimum energy would have an asymptotic Gumbel
distribution. We show that due to the hierarchical correlations, the
forward tail of the distribution of the minimum energy becomes highly
non universal, depends explicitly on the distribution of the bond
energies $\epsilon$ and is generically different from the
super-exponential forward tail of the Gumbel distribution. The
consequence of these results to the persistence of hierarchically
correlated random variables is discussed and the persistence is also
shown to be generically anomalous.}

\noindent

\date{23 March 2001}
\end{abstract}

\begin{multicols}{2}
The extreme value statistics of random variables is important in
various branches of physics, statistics, and
mathematics\cite{Gumbel,Galambos,Berman}. For example, in the context
of disordered systems, the thermodynamics at low temperatures is
governed by the statistics of the low energy states. The statistics of
extremal quantities also play important roles in binary search
problems in computer science\cite{KM}. The extreme-value statistics is
well understood when the random variables are {\em independent} and
identically distributed. In this case, depending on the distribution
of the random variable, three different universality classes of
extreme value statistics are known\cite{Berman}.  Recently there has
been an attempt to identify these different universality classes with
the different schemes of replica symmetry breaking\cite{BM}. A natural
question is: what are the universality classes when the random
variables are correlated?  This question has recently been
addressed\cite{CL,BM} and it has been conjectured that this class of
problems corresponds to the full replica symmetry
breaking\cite{BM}. To answer this important question, it would thus be
useful to derive exact results for the extreme value statistics of
correlated variables, whenever possible.

More precisely, let us consider a set of $N$ random variables $E_1$,
$E_2$, $\ldots$, $E_N$ drawn from a joint probability distribution
$p(E_1, E_2,\ldots, E_N)$. Then the minimum value, $E_{\rm min}={\rm
min}\{E_1, E_2,\ldots, E_N\}$ is also a random variable and one would
like to know its probability distribution. Let, $P_N(x)=Prob[E_{\rm
min}\ge x]$ be the cumulative distribution of the minimum. Then
clearly,
\begin{equation}
P_N(x)=\int_x^{\infty}\dots\int_x^{\infty}p(E_1,E_2,\dots,
E_N)\prod_{i=1}^{N}dE_i ,
\label{multi}
\end{equation}
since if the minimum is bigger than $x$, then each of the variables
must also be bigger than $x$.  When the variables are uncorrelated and
each is drawn from the same distribution $p(E)$, the joint
distribution factorizes, $p(E_1,E_2,\ldots, E_N)=p(E_i)\dots p(E_N)$
and from Eq. (\ref{multi}) one simply gets,
$P_N(x)=[\int_x^{\infty}p(E)dE]^N$.  If the distribution $p(E)$ is
unbounded and decays faster than a power law for large $|E|$, then one
can show that for large $N$, $P_N(x)$ approaches a scaling
form\cite{Berman}, $P_N(x)=F[(x+a_N)/b_N]$. Here $a_N$ and $b_N$ are
functions of $N$ and depend explicitly on the distribution $p(E)$, but
the scaling function $F(y)$ is independent of $p(E)$ and $N$ and has
the universal super-exponential form, $F(y)=\exp[-\exp(y)]$. As a
consequence, the distribution of the minimum $P_{\rm
min}(y)=-dF/dy=\exp[y-\exp(y)]$ has the universal Gumbel form. There
are two other known universality classes when the distribution $p(E)$
is either bounded or has algebraic tails for large $|E|$, but we will
not be concerned with these cases in this paper.

The question we focus on here is whether the Gumbel law continues to
hold if the random variables are unbounded but correlated. This
question has recently been addressed by Carpentier and Le
Doussal\cite{CL} who developed a renormalization group (RG) approach
for logarithmically correlated variables. With logarithmic
correlations they found that the cumulative distribution function
$F(y)$ behaves (up to some rescaling factors) as, $F(y)=1 -y\exp(y)$
in the {\it backward} tail region $y\to -\infty$. A pure Gumbel law
would have predicted, $F(y)=1-\exp(y)$ as $y\to -\infty$. Thus the Gumbel
law is indeed violated in this backward tail region. However, their RG
approach can not predict whether the super-exponential {\it forward}
tail of the Gumbel distribution still holds or not.  The question we
are interested in is whether strong correlations can also modify the
super-exponential {\it forward} tail of the Gumbel distribution. If
so, this has interesting consequence for the persistence of random
variables as we discuss below.

The persistence of random variables, a subject that has generated a
lot of recent interest \cite{Review}, is related to the distribution
of the minimum in a simple way. For random variables each with zero
mean, the persistence is simply the probability that all of them are
positive and is given by $P_N(0)$ in Eq. (\ref{multi}). For
independent variables, it follows trivially from Eq. (\ref{multi})
that $P_N(0)$ decays exponentially with $N$, $P_N(0)=\exp (-\theta N)$
where $\theta=-\log [\int_0^{\infty}p(E)dE]$.  For correlated
variables, this problem has been studied for many decades by applied
mathematicians who call it the `one sided barrier'
problem\cite{BL,Slepian}.  It is well known that $P_N(0)$ is hard to
compute analytically even for Gaussian correlated variables, i.e.,
when the joint distribution $p(E_1,E_2,\ldots, E_N)$ is a multivariate
Gaussian distribution\cite{BL,Slepian,Gupta}.  If the Gaussian
variables are arranged on a line and if the correlation between two
variables $E_i$ and $E_j$ decays faster than $1/|i-j|$, then $P_N(0)$
is known to decay as $P_N(0)\sim \exp(-\theta N)$ for large
$N$\cite{Slepian}, where the persistence exponent $\theta$ is
nontrivial and is known exactly only in very few special
cases\cite{BL}.  It would thus be interesting to know if strong
correlations can modify this exponential decay of the persistence for
large $N$.

In this paper, we show that the two issues, (a) the possibility of a
non-Gumbel forward tail of the distribution of the minimum and (b) the
possibility of non-exponential decay of persistence, are related to
each other for random variables that are hierarchically
correlated. The hierarchical nature of the correlation allows us to
derive exact asymptotic results for both the quantities.  Our main
results are twofold: (i) For the distribution of minimum value, we
show that the super-exponential forward tail of the Gumbel law is
violated under generic conditions and (ii) as a consequence, the
persistence is anomalous, i.e., $P_N(0)$ {\it does not} decay
exponentially under the same generic conditions.

We consider, as a model, the well studied problem of a directed polymer
on a tree.  This problem was first studied by Derrida and
Spohn\cite{DS}, who were mostly interested in the finite temperature
phase transition in this model. Here we focus explicitly on the zero
temperature properties. We consider a tree rooted at $O$ (see Fig. 1)
and a random energy $\epsilon_i$ is associated with every bond of the
tree. The variables $\epsilon_i$'s are independent and each drawn from
the same distribution $\rho(\epsilon)$. A directed polymer of size $n$
goes down from the root $O$ to any of the $2^n$ nodes at the level
$n$. Thus, there are $N=2^n$ possible paths for the polymer of size
$n$ and the energy of any of these paths is given by,
\begin{equation}
E_{\rm path}=\sum_{i\in {\rm path}}{\epsilon}_i .
\label{energy}
\end{equation}
The set of $N=2^n$ variables $E_1$, $E_2$, $\ldots$, $E_N$ are clearly
correlated in a hierarchical (i.e. ultrametric) way and the two point
correlation between the energies of any two paths is proportional to
the number of bonds they share. We would then like to know the distribution
of the minimum energy.
\begin{figure}
\narrowtext\centerline{\epsfxsize\columnwidth \epsfbox{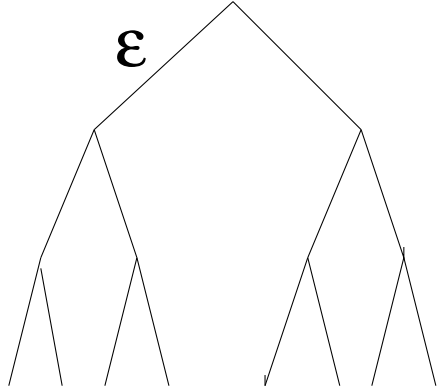}}
\caption{Each bond of a Cayley tree has an energy $\epsilon_i$.}
\label{fig1}
\end{figure}

Clearly, $P_N(x)=Prob[E_{\rm min}\ge x]$ is also the probability that
all the $N$ paths up to the $n$-th level have energies $\ge x$. Since
$N=2^n$, let us write, for convenience, $R_n(x)=P_N(x)$. It is easy to
see that $R_n(x)$ satisfies the recursion relation,
\begin{equation}
R_{n+1}(x)={\left[ \int_{-\infty}^{\infty} d\epsilon \rho(\epsilon)
R_n(x-\epsilon)\right]}^2,
\label{recur1}
\end{equation}
with the initial condition, $R_0(x)=\theta(-x)$ where $\theta(x)$ is
the usual Heaviside step function.  This relation is derived by
considering various possibilities for the energies of the two bonds
emerging from the root $O$ and taking into account that the two
subsequent daughter trees are statistically independent. The
Eq. (\ref{recur1}) was studied in detail in Ref. \cite{MK} for several
distributions $\rho(\epsilon)$'s with non negative support. In
particular, for the bivariate distribution,
$\rho(\epsilon)=p\delta(\epsilon-1)+(1-p)\delta (\epsilon)$, the
solution of Eq. (\ref{recur1}) was shown to undergo a depinning phase
transition at $p_c=1/2$\cite{MK}. Since in this paper we are mostly
interested in the persistence of the $E_i$ variables, we restrict
ourselves subsequently only to symmetric distributions
$\rho(\epsilon)$ with zero mean. Defining $R_n(x)=Q_n^2(x)$,
Eq. (\ref{recur1}) can be recast into,
\begin{equation}
Q_{n+1}(x)=\int_{-\infty}^{\infty}d\epsilon \rho(\epsilon)
Q_n^2(x-\epsilon),
\label{recur2}
\end{equation}
with the initial condition, $Q_0(x)=\theta(-x)$ and the boundary
conditions, $Q_n(x)\to 0$ as $x\to \infty$ and $Q_n(x)\to 1$ as $x\to
-\infty$.

The Eq. (\ref{recur2}) is known\cite{MK} to admit a traveling front
solution, $Q_n(x)=q(x+vn)$ where the front propagates in the negative
$x$ direction with a constant velocity $v$ as $n$ increases (see
Fig. 2). Substituting $Q_n(x)=q(x+vn)$ in Eq. (\ref{recur2}), we get
\begin{equation}
q(y)=\int_{-\infty}^{\infty}d\epsilon \rho(\epsilon)
q^2(y-v-\epsilon),
\label{recur3}
\end{equation}
with the boundary conditions, $q(y)\to 1$ as $y\to -\infty$ and
$q(y)\to 0$ as $y\to \infty$, with the front located around $y=0$.
The velocity $v$ can then be determined exactly by analyzing the
backward tail region, $y\to -\infty$ of the function $q(y)$. In this
regime, substituting $q(y)=1-g(y)$ in Eq. (\ref{recur3}) and
neglecting the terms of $O(g^2)$, we find that the resulting linear
equation admits an exponential solution, $g(y)=\alpha \exp(\lambda y)$
with $\alpha>0$ provided $v$ is related to $\lambda$ via the
dispersion relation,
\begin{equation}
v= {1\over {\lambda}}\log\left[ 2\int_{-\infty}^{\infty}d\epsilon
\rho(\epsilon)e^{-\lambda \epsilon}\right].
\label{vel}
\end{equation}
For generic distributions $\rho(\epsilon)$, the function $v_{\lambda}$
has a unique minimum at $\lambda=\lambda^*$ and by the general
velocity selection principle\cite{VS}, this minimum velocity,
$v_{\lambda^*}$ is selected by the front\cite{DS,MK}.

Thus the cumulative distribution of the minimum energy approaches a
scaling form for large $N$, $P_N(x)=R_n(x)=Q_n^2(x)\to q^2[x+{
{v_{\lambda^*}}\over {\log 2} }\log N]$, where the function $q(y)$ is
given by the solution of Eq. (\ref{recur3}) and $v_{\lambda^*}$ is
determined by minimizing Eq. (\ref{vel}). The question we are
interested in is: what is the asymptotic form of $q(y)$ for large $y$?
We show below that that for any bounded distribution $\rho(\epsilon)$,
the function $q(y)$ for large $y$ indeed has the Gumbel shape,
$q(y)\to \exp[-c_1\exp (c_2 y)]$ where $c_1$ and $c_2$ are positive
constants.  On the other hand, for unbounded distributions
$\rho(\epsilon)$, the Gumbel law breaks down and asymptotic forward
tail of $q(y)$ is nonuniversal and is determined explicitly by the
distribution $\rho(\epsilon)$.  For example, for the exponential
distribution, $\rho(\epsilon)=\exp[-|\epsilon|]/2$, we find exactly
$q(y)\to \exp(-y)$ for large $y$.  For a generic unbounded distribution,
one can prove a lower bound, $q(y)> f(y)$ for large $y$, where
$f(y)=\int_y^{\infty} \rho(\epsilon)d\epsilon$.
\begin{figure}
\narrowtext\centerline{\epsfxsize\columnwidth \epsfbox{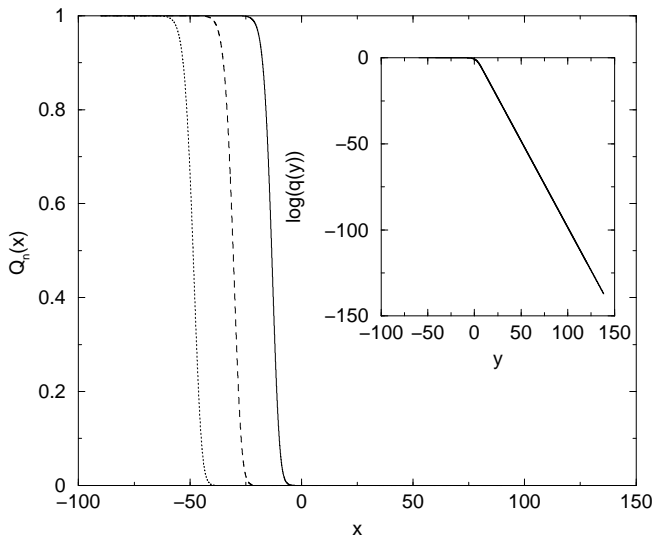}}
\caption{The traveling front for the function $Q_n(x)$ for $n=10$ (the
solid line), $20$ (the dashed line) and $30$ (the dotted line) for
exponential distribution $\rho(\epsilon)=\exp[-|\epsilon|]/2$.  In the
inset, we plot the logarithm of the collapsed scaling function
$q(x+vn)$ (for different $n$).  The scaling function $q(y)$ evidently
has an exponential tail for large $y$.}
\label{fig2}
\end{figure}

We first focus on the unbounded distributions $\rho(\epsilon)$. Let us
first consider the exponential distribution,
$\rho(\epsilon)=\exp(-|\epsilon|)/2$. In this case, by first making a
change of variable $\epsilon\to y-v-\epsilon$ inside the integrand on
the right hand of side of Eq. (\ref{recur3}) and then differentiating
twice the resulting equation, we get
\begin{equation}
{ {d^2q}\over {dy^2}}=q(y)-q^2(y-v).
\label{qy}
\end{equation}
For large $y$, clearly the nonlinear term is negligible since $q(y)$
is small.  Using the boundary condition $q(y)\to 0$ as $y\to \infty$
we then get, $q(y)\to A \exp(-y)$ for large $y$ where $A$ is a
constant. Thus we get an exponential forward tail instead of the
standard super-exponential forward tail of the Gumbel distribution.
Note that the velocity $v$ is determined, as before, from the $y\to
-\infty$ tail where $q(y)=1-\alpha e^{\lambda y}$ and Eq. (\ref{qy})
gives, $v_{\lambda}=\log[2/(1-\lambda^2)]/\lambda$ with $0<\lambda<1
$, in accordance with the general formula in Eq. (\ref{vel}). The
function $v_{\lambda}$ has a unique minimum at
$\lambda^*=0.603582\dots$ and the chosen front velocity is then,
$v_{\lambda^*}=1.89899\dots$. In Fig. (2), we show that $Q_n(x)$
indeed approaches the scaling form, $Q_n(x)\to q(x+v_{\lambda^*}n)$
and the tail of the scaling function is given by, $q(y)\sim \exp(-y)$
(see the inset of Fig. (2)) as predicted analytically.

For a generic unbounded distribution it is difficult to derive exact
results. However, one can easily derive a lower bound for $q(y)$. From
Eq. (\ref{recur3}), it is clear that $q(y)\ge
\int_{y-v}^{\infty}d\epsilon \rho(\epsilon) q^2(y-v-\epsilon)$.  This
follows since the integrand on the right hand hand side of
Eq. (\ref{recur3}) is always positive. Since the function $q(y)$
saturates to $1$ very quickly for negative $y$, we can replace
$q^2(y-v-\epsilon)$ by $1$ on the right hand side of the above lower
bound. This gives, for large $y$, $q(y)\ge f(y)$ where
$f(y)=\int_y^{\infty}\rho(\epsilon)d\epsilon $. For example, for the
Gaussian distribution, $\rho(\epsilon)=e^{-\epsilon^2/2}/{\sqrt
{2\pi}}$, this result indicates that $q(y)$ should decay at most as
fast as $f(y)={\rm erfc}(y/{\sqrt 2})$. Thus, for generic unbounded
distributions, the forward tail of the function $q(y)$ for large $y$
is highly nonuniversal and is generally different from the
super-exponential forward tail as in the Gumbel distribution.

\begin{figure}
\narrowtext\centerline{\epsfxsize\columnwidth \epsfbox{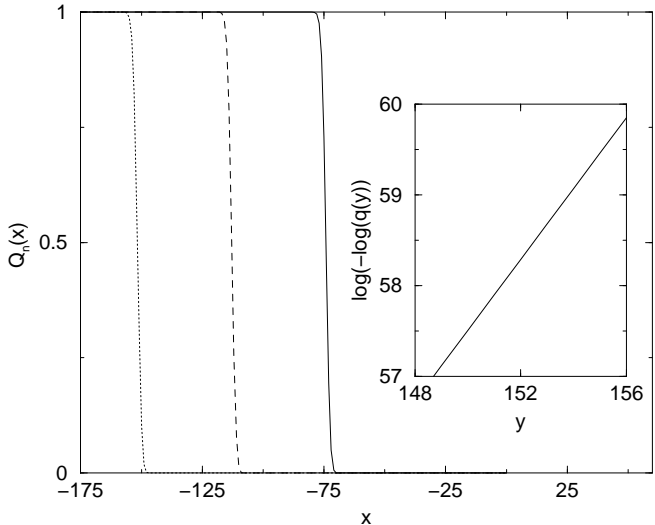}}
\caption{The traveling front for the function $Q_n(x)$ for $n=100$
(the solid line), $150$ (the dashed line) and $200$ (the dotted line)
for the distribution $\rho(\epsilon)=
a\delta(\epsilon+1)+a\delta(\epsilon-1)+(1-2a)\delta(\epsilon)$ with
$a = 1/4$.  In the inset, we plot $\log(-\log(q(y)))$ against $y$ 
which clearly shows the super-exponential decay of
the scaling function $q(y)$ for large $y$.}
\label{fig3}
\end{figure}

Next we consider the bounded distributions $\rho(\epsilon)$. The lower
bound discussed in the previous paragraph continues to hold for
bounded distributions as well, though for large $y$ it trivially
becomes zero for distributions with an upper cutoff. To obtain more
precisely the behavior of $q(y)$ as $y\to \infty$, we first consider a
specific example,
$\rho(\epsilon)=a\delta(\epsilon+1)+a\delta(\epsilon-1)
+(1-2a)\delta(\epsilon)$ with $0<a<1/2$. The Eq. (\ref{recur3}) then
becomes,
\begin{eqnarray}
q(y)&=&a\,q^2(y-v-1)+a\, q^2(y-v+1) \nonumber \\ &+&(1-2a)q^2(y-v),
\label{qy1}
\end{eqnarray}
where the velocity $v=v_{\lambda^*}$ is obtained by minimizing
Eq. (\ref{vel}) with respect to $\lambda$. In this particular case, we
get from Eq. (\ref{vel}),
\begin{equation}
v_{\lambda}={1\over {\lambda}} \log\left[4a\cosh
(\lambda)+2(1-2a)\right],
\label{velb}
\end{equation}
which has a unique minimum at $\lambda=\lambda^*(a)$ for all
$0<a<1/2$.  We then need to analyze the large $y$ behavior of $q(y)$
in Eq. (\ref{qy1}) with $v=v_{\lambda^*}$.  Note that as one increases
$y$ from $-\infty$, $q(y)$ remains approximately $1$ up to the back
edge of the front at $y=0$ and then starts decreasing to $0$ as $y$
increases beyond $0$.  The idea would be to determine $q(y)$ for a
fixed large $y$ by iterating Eq. (\ref{qy1}) backwards in $y$ till we
reach the back edge of the front at $y=0$ where $q(y)\approx
1$. Anticipating a super-exponential decay of $q(y)$ for large $y$,
one can neglect the second and the third term on the right hand side
of Eq. (\ref{qy1}) and iterate the equation retaining only the first
term. Iterating $m$ times backward we get,
\begin{equation}
q(y)\approx a^{2^m-1}{\left[q\left(y-m(v+1)\right)\right]}^{2^m}.
\label{qyap}
\end{equation}
How many iterations do we need to reach $0$ starting from a fixed
large $y$? Clearly the required value of $m$ is given by, $m=y/(v+1)$
so that the argument of the function on the right hand side of
Eq. (\ref{qyap}) becomes $0$. Using $q(0)\approx 1$, we get from
Eq. (\ref{qyap}) the large $y$ behavior,
\begin{equation}
q(y)\approx a^{2^{y/(v+1)}},
\label{sexp}
\end{equation}
confirming the super-exponential forward tail of $q(y)$ and also
justifying, a posteriori, the neglect of the second and the third term
in the iteration of Eq. (\ref{qy1}). We have verified the analytical prediction
in Eq. (\ref{sexp}) by direct numerical integration of
Eq. (\ref{recur2}) with $\rho(\epsilon)=a\delta(\epsilon+1)+a\delta(\epsilon-1)
+(1-2a)\delta(\epsilon)$ for $a=1/4$ (see Fig. (\ref{fig3})).

The argument leading to the result in Eq. (\ref{sexp}) above uses only
the fact that distribution $\rho(\epsilon)$ has an upper cutoff at
$\epsilon=1$. Thus we expect that $q(y)$ will always have a
super-exponential forward tail as long as the distribution
$\rho(\epsilon)$ is bounded with an upper cutoff $\Lambda$.  Let us
consider another example of a bounded distribution, namely the uniform
distribution,
$\rho(\epsilon)=[\theta(\epsilon+1)-\theta(\epsilon-1)]/2$. In this
case, we get from Eq. (\ref{recur3}),
\begin{equation}
q(y)={1\over {2}}\int_{y-v-1}^{y-v+1}dz q^2(z).
\label{qy2}
\end{equation}
Differentiating the Eq. (\ref{qy2}) with respect to $y$ yields,
\begin{equation}
{ {dq}\over {dy}}={1\over {2}}q^2(y-v+1)-{1\over {2}}q^2(y-v-1).
\label{qy3}
\end{equation}
Again we anticipate that $q(y)$ will have a super-exponential tail for
large $y$.  If so, one can make the approximation, $\log(-dq/dy)
\approx \log(q(y))$.  Using this in Eq. (\ref{qy3}) we iterate the
equation backwards as before after dropping the first term on the
right hand side of Eq. (\ref{qy3}). Using the same line of arguments
used in the previous paragraph, we finally get a super-exponential
tail for large $y$ as before,
\begin{equation}
q(y)\approx 2^{-2^{y/(v+1)}}.
\label{sexp1}
\end{equation}
Note that the velocity $v$ in Eq. (\ref{sexp1}) has to be determined
by minimizing Eq. (\ref{vel}) with a uniform distribution.  Thus, in
general, for any bounded distribution, we expect that for large $y$
\begin{equation}
q(y)\approx \exp\left[-c\, 2^{y/(v+1)}\right],
\label{sexp2}
\end{equation}
where the constants $c$ and $v$ depend explicitly on the distribution
$\rho(\epsilon)$.  

\begin{figure}
\narrowtext\centerline{\epsfxsize\columnwidth \epsfbox{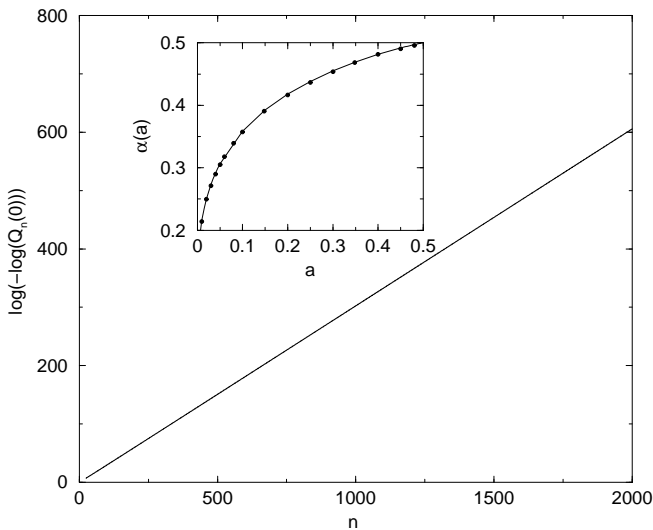}}
\caption{ The function $\log(-\log(Q_n(0)))$, obtained
from the numerical integration of Eq. (\ref{recur2}), is plotted against $n$
for the distribution $\rho(\epsilon)=a\delta(\epsilon+1)+
a\delta(\epsilon-1)+(1-2a)\delta(\epsilon)$ with $a=1/4$. The linear increase
with $n$ confirms the stretched exponential decay of $P_N(0)=Q_n^2(0)$
for large $N=2^n$. In the inset is shown the value of
$\alpha(a)$ calculated analytically (solid line) with that measured
numerically by direct integration of Eq. (\ref{recur2}) (circles).}
\label{fig4}
\end{figure}

Having established the forward tail behavior of
$R_n(x)=q^2(x+vn)$, we now turn to the persistence. The persistence is
simply given by $P_N(0)=R_n(0)=Q_n^2(0)=q^2(vn)$, where $N=2^n$.  Thus for
large $N$ or equivalently for large $n$, the asymptotic behavior of
persistence $P_N(0)=q^2(vn)$ is governed by the forward tail of the
function $q(y)$ for large $y$.  Let us first consider the bounded
distributions. In this case, using the result from Eq.  (\ref{sexp2})
for $q(y)$, we get the following exact result for persistence for
large $N=2^n$,
\begin{equation}
P_N(0)=Q_n^2(0)=q^2(vn)\approx \exp\left[-2c N^{\alpha}\right],
\label{se}
\end{equation}
where $\alpha=v/(v+1)$ and $v$ is determined by minimizing
Eq. (\ref{vel}). Thus the persistence has an anomalous stretched
exponential decay for large $N$ instead of the standard exponential
decay. We have verified this analytical prediction by numerically
integrating Eq. (\ref{recur2}) for different bounded distributions. In
Fig. (4), we show the result for the
distribution $\rho(\epsilon)=a\delta(\epsilon+1)+
a\delta(\epsilon-1)+(1-2a)\delta(\epsilon)$. In this case,
$\alpha=v/(v+1)$ where $v$ is the minimum value of the dispersion relation
in Eq. (\ref{velb}) and is clearly a continuous function of the parameter $a$.
In the inset of Fig. (\ref{fig4}), we compare the analytical
prediction for the exponent $\alpha(a)=v/(1+v)$ with that obtained from
the numerical integration for various values of $a$. The agreement is
evidently very good.

\begin{figure}
\narrowtext\centerline{\epsfxsize\columnwidth \epsfbox{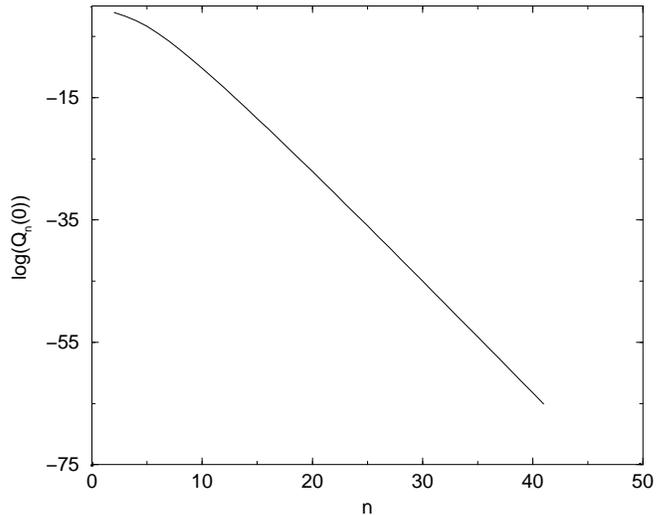}}
\caption{ The function $\log(Q_n(0)))$, obtained
from the numerical integration of Eq. (\ref{recur2}), is plotted against $n$
for the distribution $\rho(\epsilon)=\exp(-|\epsilon|)/2$.
The linear decrease
with $n$ confirms the power law decay of $P_N(0)=Q_n^2(0)$
for large $N=2^n$ with the exponent $\beta=2v/\log 2$ where $v \approx 1.89$.}
\label{fig5}
\end{figure}

We next consider the unbounded distributions such as
$\rho(\epsilon)=\exp[-|\epsilon|]/2$. For this exponential
distribution, using the asymptotic behavior $q(y)=A\exp(-y)$, we find
that for large $N$,
\begin{equation}
P_N(0)=Q_n^2(0)=q^2(vn)\sim \exp[-2vn] \sim N^{-\beta},
\label{pl}
\end{equation}
where $\beta=2v/\log 2$ with $v$, as usual, determined via minimizing
Eq. (\ref{vel}).  Thus in this case, persistence again decays
anomalously but now as a power law with a nonuniversal exponent $\beta$. 
Again we verified this analytical prediction 
numerically by directly integrating
Eq. (\ref{recur2}) with the exponential distribution (see Fig. (\ref{fig5})).
For generic unbounded distributions, using the lower bound $q(y)\ge f(y)$
for large $y$ where $f(y)=\int_y^{\infty}\rho(\epsilon)d\epsilon$, we
get $P_N(0)=q^2(nv)\ge f^2[{v\log N}/\log 2]$, again highly anomalous.

In summary, we have investigated in detail the distribution of the
minimum energy of a directed polymer on a Cayley tree.  We have shown
that the hierarchical correlations between the energies of different
paths have a considerable effect on the distribution of minimum energy
depending on the distribution of bond energies $\rho(\epsilon)$.  In
the case of bounded distributions $\rho(\epsilon)$ of the bond
energies, we have shown that the forward tail has a super-exponential
tail as in the Gumbel distribution. However, for unbounded
distributions $\rho(\epsilon)$ the forward tail is highly non
universal and depends explicitly on the distribution
$\rho(\epsilon)$. This rich behavior of the forward tail of the
minimum energy distribution is shown to lead to a variety of anomalous
behavior for the persistence probability $P_N(0)$, ranging from a
stretched exponential decay for bounded distributions $\rho(\epsilon)$
to power law decay when $\rho(\epsilon)$ is exponential.

\end{multicols}
\end{document}